\documentclass[conference]{IEEEtran}
\usepackage{amssymb}
\usepackage{graphicx}
\usepackage{amsmath,amsthm,amssymb}
\usepackage{amsfonts}
\usepackage{latexsym,psfrag}
\usepackage{mdwlist}
\DeclareMathOperator{\rank}{rank}
\DeclareMathOperator{\score}{score}
\newtheorem{thm}{Theorem}

\newcommand{\gc}[2]{{{#1}\brack{#2}}}
\ifCLASSINFOpdf
\else
\fi
\usepackage{algorithmic}
\usepackage{algorithm}
\begin{document}
\title{Projective Space Codes for the Injection Metric}
\author{\IEEEauthorblockN{Azadeh Khaleghi,
Frank R. Kschischang}
\IEEEauthorblockN{Department of Electrical and Computer Engineering, University of Toronto, Canada\\ Email: \{azalea,frank\}@comm.utoronto.ca}
}
\maketitle
\begin{abstract}

In the context of error control in random linear network coding,
it is useful to construct codes that comprise well-separated collections
of subspaces of a vector space over a finite field.  In this paper,
the metric used is the so-called ``injection distance,'' introduced
by Silva and Kschischang. A Gilbert-Varshamov bound for such codes is
derived.  Using the code-construction framework of
Etzion and Silberstein, new non-constant-dimension codes are
constructed;  these codes contain more codewords than comparable
codes designed for the subspace metric.
\end{abstract}

\IEEEpeerreviewmaketitle
\section{Introduction}
The problem of error-correction in random network coding has recently become an active area of research \cite{N5, N7, N13, N12, KK, SKK}. The main motivation for this problem is the phenomenon of error-propagation in the network. Since the received packets are random linear combinations of packets inserted at intermediate nodes, the system is very sensitive to transmission errors. Due to the vector-space preserving nature of random linear network coding, it has been shown that codes constructed in the projective space are suitable for error-correction for network coding. 

Our focus in this paper is on construction of codes in the projective space for adversarial error-correction in random network coding. As shown in \cite{metrics} a suitable measure of distance between subspaces for an adversarial error-control model is given by 
\begin{eqnarray}
d_I(U,V) = \max\{\dim U, \dim V\} - \dim(U \cap V) \nonumber,
\end{eqnarray}
a measure known as the ``\textit{injection metric}". 
This choice of distance metric is the main parameter that distinguishes our work from the existing literature on (subspace) codes constructed for random linear network coding. All existing bounds and constructions are based on a metric known as the subspace distance $d_S$ originally introduced by K\"otter and Kschischang in \cite{KK}. In the special case where codes are contained in the Grassmannian, codes designed for $d_I$ coincide with those designed for $d_S$. However, as shown in \cite{metrics}, in general non-constant dimension codes designed for $d_I$ may have higher rates than those designed for $d_S$.

In this paper we present a construction of a class of codes in the projective space for the injection distance. This construction is motivated by the work of Etzion and Silberstein in \cite{ES}, with the main difference that the construction in \cite{ES} is based on $d_S$.

In Section~\ref{pre}, we present a brief overview of the projective space and the Grassmannian, as well as rank-metric codes. We also briefly review Etzion and Silberstein's ``Ferrers diagram lifted rank-metric codes", as our work in Section~\ref{const} is related to this construction. In Section~\ref{gv}, we present a Gilbert-Varshamov-type bound on the size of codes of a certain minimum injection distance in the projective space. As we are precluded by space in this paper, we present this theorem without proof. In Section~\ref{fdrm} we present a construction for the Ferrers diagram rank-metric codes as subcodes of linear MRD codes. In Sections~\ref{const} and ~\ref{alg}, we provide an algorithm for the construction of a class of non-constant-dimension codes in the projective space designed for $d_I$. Finally, in Section~\ref{params} we present our numerical results. As shown in this section our construction results in codes of slightly higher rates than the codes of \cite{ES}.

\section{Preliminaries}\label{pre}
\subsection{Notation}
Let $q \geq 2$ be a power of a prime. In this paper, all vectors and matrices are defined over the finite field $\mathbb F_q$, unless otherwise mentioned. We denote by $\mathbb F_q^{m \times n}$, the set of all $m \times n$ matrices over $\mathbb F_q$. If $v$ is a vector then the $i^{th}$ entry of $v$ is denoted by $v_i$. We denote the logical complement of a binary vector $v = (v_1,v_2,\cdots,v_n)$ by $\bar v = (\bar v_1,\bar v_2,\cdots,\bar v_n)$. The number of non-zero elements of $v$ is denoted by $wt(v)$. We define the support set of a vector $v$, denoted $\operatorname{supp}(v)$ to be the set of indices corresponding to the non-zero entries of $v$. Let $x$ and $y$ be two binary vectors of the same length. We denote the number of $1 \rightarrow 0$ transitions from $x$ to $y$ by $N(x,y)$, their Hamming distance by $d_H(x,y)$ and the logical AND operation between $x$ and $y$ by $\wedge$. If $X$ is a matrix then the rank of $X$ is denoted by $\rank X$ and its row space is denoted by $\left < X \right >$. Let $n > 0$ be an integer. We denote by $[n]$ the set of all positive integers less than or equal to $n$, i.e. $[n] = \{0,1,2,\cdots, n\}$.
\subsection{Rank-Metric Codes}
Let $X$ and $Y$ be two matrices in $\mathbb F_q^{m \times n}$. The \textit{rank distance} between $X$ and $Y$, denoted $d_R(X,Y)$ is defined as $d_R(X,Y) \triangleq \rank(Y-X)$. As shown in \cite{gab} the rank distance is indeed a metric. Let $\mathbb F_q$ be a base field and $\mathbb F_{q^m}$ with $m \geq 1$ be an extension of $\mathbb F_q$. The rank of a vector $v=(v_1,v_2,\cdots,v_n) \in (\mathbb F_{q^m})^n$ is the rank of the $m \times n$ matrix obtained by expanding each entry of $v$ to an $m \times 1$ column vector over $\mathbb F_q$. A code $C_\mathcal R$ is a rank-metric code over $\mathbb F_{q^m}$ of minimum distance $d$, if $\mathcal C_{\mathcal R} \subseteq (\mathbb F_{q^m})^n$ and for all $X, \; Y \; \in C_{\mathcal R} \; d_R(X,Y) \geq d$. As shown in \cite{gab} $C_{\mathcal R}$ must satisfy $ \log_q \left|\mathcal C_{\mathcal R}\right| \leq \max  \{m,n\}(\min\{m,n\}-d+1)$, and rank metric codes achieving this bound with equality are said to be Maximum Rank Distance (MRD) codes. Gabidulin codes,  presented by Gabidulin \cite{gab} are an extensive class of MRD codes, which are the analogs of the generalized Reed-Solomon codes designed for the rank metric. Efficient polynomial-time decoding algorithms exist that correct errors of rank up to $\left \lfloor \displaystyle{\frac{d-1}{2}}\right \rfloor$. See for example \cite{gabdec2, gabdec3, gabdec4}.

\subsection{Projective Space}
Let $V$ be an $n$-dimensional vector space over the finite field $\mathbb F_q$ of order $q$. For a non-negative integer $k \leq n$ denote by $\mathcal{G}(n,k)$ the set of all $k$-dimensional subspaces of $V$. This set is known as a \textit{Grassmannian} and its cardinality is given by the $q$-ary \textit{Gaussian coefficient} defined as $\displaystyle \gc{n}{k}_q \triangleq \frac{(q^n-1)(q^{n-1}-1)\cdots(q^{n-k+1}-1)}{(q^k-1)(q^{k-1}-1)\cdots(q-1)}$. The set of all subspaces of $V$ form a \textit{projective space} $\mathcal P_q^n$ of order $n$ over $\mathbb F_q$. Thus $\mathcal P_q^n$ can be viewed as a union of the Grassmannians for all $k\leq n$, i.e. $\mathcal P_q^n = \displaystyle{\bigcup_{k=0}^n} \mathcal{G}(n,k)$. A code $\mathcal C$ is an $(n,M,d)_{d_S}$ code in $\mathcal P_q^n$ if $\left | \mathcal C \right | = M$  and for all $U, \; V \in \mathcal C, \;d_S(U,V) \geq d$. Similarly, a code $\mathcal C \subseteq \mathcal P_q^n$ is an $(n,M,d)_{d_I}$ code if $\left | \mathcal C \right| = M$  and for all $U, \; V \in \mathcal C, \;d_I(U,V) \geq d $. A code $\mathcal C$ is an $(n,M,d,k)$ constant-dimension code if $\mathcal C \subseteq \mathcal G(n,k)$ for some $k \in [n]$. Since in this case $d_I$ and $d_S$ are equal up to scale, there is no need to distinguish between $(n,M,d,k)_{d_I}$ and $(n,M,d,k)_{d_S}$ codes.

\subsection{Ferrers Diagram Lifted Rank-Metric Codes}
In this section we review the code construction of \cite{ES} with a slightly different notation. The key idea in this construction is the observation that every $k$-dimensional vector space $V$ in $\mathcal P_q^n$ arises uniquely as the row space of a $k \times n$ matrix in Reduced Row Echelon Form (RREF). Let $V$ be a vector space in $\mathcal P_q^n$and let $E(V)$ be its corresponding generator matrix in RREF. We define the \textit{profile vector} of $V$ denoted $p(V)$, to be a binary vector of length $n$ whose non-zero elements appear \textit{only} in positions where $E(V)$ has a leading $1$. Consider an equivalence relation $  \sim $ on $\mathcal P_q^n$ where, 
\begin{eqnarray}
\forall \; V_1, \;V_2 \in \mathcal P_q^n, \; V_1 \sim V_2 \leftrightarrow p(V_1) = p(V_2). \label{rel}
\end{eqnarray}
This relation partitions $\mathcal P_q^n$ into equivalence classes, where $V_1$ and $V_2$ belong to the same class provided that they are identified by the same profile vector. Let $\Gamma$ denote the set of all equivalence classes generated in $\mathcal P_q^n$ according to (\ref{rel}). Consider an equivalence class $\gamma \in \Gamma$ with a profile vector $v$ of length $n$ and weight $k$. We define the profile matrix $P_M(v)$ to be a $k \times n$ matrix in RREF where the leading coefficients of its rows appear in columns indexed by $\operatorname{supp}(v)$, and has $\bullet$'s in all its entries which are not required to be terminal zeros or leading ones. For example if $p=(0,1,0,1,1,0,0)$ then $P_M(v) = \left[ \begin {array}{ccccccc} 0&1&\bullet&0&0&\bullet&\bullet\\0&0&0&1&0&\bullet&\bullet\\0&0&0&0&1&\bullet&\bullet\end {array} \right]$. Notice that the generator matrices in RREF of the elements of $\gamma$ differ only in entries of $P_M(v)$ marked as $\bullet$'s. Let $\eta$ denote the number of columns of $P_M(v)$ which contain at least a single $\bullet$. Let $S(v)$ be the $k \times \eta$ sub-matrix of $P_M(v)$ composed of all such columns of $P_M(v)$. A code is an $[S,\kappa, \delta]$ Ferrers diagram rank-metric code if it forms a rank-metric code with dimension $\kappa$ and minimum rank-distance $\delta$, all of whose codewords are $m \times \eta$ matrices with zeros in all their entries where $S(v)$ has zeros.

In the construction presented in \cite {ES}, a set $\Omega \subseteq \Gamma$ is constructed in such a way that for all $\gamma_1, \; \gamma_2 \in \Omega$ with $\gamma_1 \neq \gamma_2$ and for all $V_1 \in \gamma_1, V_2 \in \gamma_2, \; d_S(V_1,V_2) \geq d$. By Lemma 2 in \cite{ES}, this is possible by selecting the profile vectors of the equivalence classes according to a binary code of minimum Hamming distance $d$. Then within each class $\gamma \in \Omega$, a Ferrers diagram rank-metric code is used to ensure that for all $V_1, \; V_2 \in \gamma, d_S(V_1,V_2) \geq d$. Finally $\mathcal C = \{V \in \gamma| \gamma \in \Omega\}$. 

\section{A Gilbert-Varshamov-Type Bound on the Size of Codes in the Projective Space}\label{gv}
Let $V$ be a $k$-dimensional vector space in $\mathcal P_q^n$. We define $S_t(V)$ to be the set of all vector spaces in $\mathcal P_q^n$ at an injection distance at most $t$ from $V$. i.e.
\begin{eqnarray}
 S_t(V) = \{W \in \mathcal {P}^n | d_I(V,W) \leq t\} \nonumber
\end{eqnarray}
We may view $S_t(V)$ as a hypothetical \textit{sphere} of radius $t$ centered at $V$. In Theorem~\ref{size} we give the cardinality of $S_t(V)$ centered at some $k$-dimensional vector space with $k\leq n$. Since the projective space is non-homogeneous, the size of $S_t(V)$ does not depend merely on its radius, but also on the dimension of its center. In other words for two vector spaces $V_1$ and $V_2$ with $\dim V_1 \neq \dim V_2$, we have $\left | S_t(V_1) \right | \neq  \left | S_t(V_2) \right |$. 
\begin{thm}\label{size}
Let $V$ be a $k$-dimensional vector space in $\mathcal P_q^n$, with $k \leq n$, and let $\mathcal N(k,t)$ denote the cardinality of $S_t(V)$. Then,
\begin{eqnarray}
\mathcal N(k,t)& = &\displaystyle \sum_{r=0}^t q^{r^2}\gc{k}{r}_q\gc{n-k}{r}_q+ \nonumber \\
&\displaystyle\sum_{j=1}^{r}& q^{r(r-j)}\left(\gc{k}{r}_q\gc{n-k}{r-j}_q+\gc{n-k}{r}_q\gc{k}{r-j}_q\right) \nonumber
\end{eqnarray}
\end{thm}
Using Theorem~\ref{size}, and following an approach similar to that of Etzion and Vardy in \cite{tol}, we obtain the following generalized Gilbert-Varshamov-type bound on the size of codes in the projective space.
\begin{thm}\label{GV}
Let $\mathcal A_q(n,d)$ denote the maximum number of codewords in an $(n,M,d)$ code in $\mathcal P_q^n$. Then,
\begin{eqnarray}
\displaystyle \mathcal A_q(n,d) \geq \displaystyle \frac{\displaystyle \left | \mathcal P_q^n \right |^2}{\displaystyle \sum _{k=0}^n \displaystyle \gc{n}{k}\displaystyle \mathcal N (k,d-1) } \nonumber
\end{eqnarray}
\end{thm}
\section{Ferrers Diagram Rank-Metric Code Construction}\label{fdrm}
Let $v$ be a binary vector of length $n$ and weight $m$. Let $C_\mathcal F $ be an $[S(v),\kappa, \delta]$ Ferrers diagram rank-metric code that fits $S(v)$, i.e. every codeword in $C_\mathcal F $ has zeros in all its entries where $S(v)$ has zeros. We may view $C_\mathcal F$ as a subcode of a linear rank-metric code $C$ of minimum rank-distance $d_R(C) \geq \delta$, with a further set of linear constraints ensuring that $C_\mathcal F $ fits $S(v)$. 

In Theorem~\ref{kappa} we provide a lower bound on the dimension $\kappa$ of the largest $[S(v),\kappa, \delta]$ Ferrers diagram rank-metric code obtained as a subcode of a linear MRD code. 
\begin{thm}\label{kappa}
Let $v$ be a binary vector of weight $m$ and let $S(v)$ be the $m \times \eta$ sub-matrix of $P_M(v)$ composed of all the columns of $P_M(v)$ that contain at least a single $\bullet$. Assume that $S(v)$ contains a total of $w$ $\bullet$'s. Consider the dimension $\kappa$ of the largest $[S, \kappa, \delta]$ Ferrer's diagram rank-metric code  $C_\mathcal F$. We have, $\kappa \geq w - \max\{m,\eta\}(\delta -1) \nonumber$.
%Let $v$ be a binary vector of weight $m$ and assume that $S(v)$ contains a total of $w$ $\bullet$'s. Let $C_\mathcal F$ be the largest $[S, \kappa, \delta]$ Ferrers diagram rank-metric code obtained as a subcode of a linear MRD code $C$ with $d_R \geq \delta$. We have,
%\begin{eqnarray}
%\kappa \geq w - \max\{m,\eta\}(\delta -1) \nonumber
%\end{eqnarray}
\end{thm}
\begin{proof}
Let $ V = \mathbb F_q^{m \times \eta}$. Note that $\mathbb F_q^{m \times \eta}$ is an $m\eta$-dimensional vector space over $\mathbb F_q$. Let $C$ be a linear MRD code with $d_R(C) \geq \delta$. This code is a $k$-dimensional subspace of $\mathbb F_q^{m \times \eta}$ with $k = \max\{m,\eta\}(\min \{m,\eta\} - \delta +1)$. There exists a linear transformation $\Phi : V \longrightarrow V/C$ with $\ker \Phi = C$, and by the First Isomorphism Theorem $\dim V/C = m\eta - k$. Let $A = \{(i,j)|S(v)_{ij} = 0\}$ be the set of $(i,j)$ indices where $S(v)$ has zeros, and note that $\left | A\right| = m\eta-w$. Let $f: V \longrightarrow \mathbb F_q^{m\eta-w}$ such that $f(x)=(x_{ij}), \; (i,j) \in A$. Now any subcode $C'$ of $C$ satisfying $f(c) = 0 \; \forall c \in C'$ is an $[S(v),\kappa, \delta]$ Ferrers diagram rank-metric code. Let $C_\mathcal F$ be \textit{the largest} such subcode of $C$. Define a linear transformation $\Phi': V \longrightarrow \mathbb V/C \times \mathbb F_q^{m\eta-w}$, by which $x \mapsto (\Phi(x), f(x))$. Now by construction $\ker \Phi' = C_\mathcal F$. Noting that $\Phi'(V) \subseteq V/C \times \mathbb F_q^{m\eta-w}$ we have $\dim \Phi'(V) \leq 2m\eta - k -w$, and by the rank-nullity theorem we obtain $ \dim C_\mathcal F \geq w+k-m\eta = w-\max\{m,\eta\}(\delta-1)$, and the theorem follows.
\end{proof}

As an example, given a profile vector $v$ of length $n$, with $wt(v)=m$ we may construct an $[S(v),\kappa,d]$ code by taking a Gabidulin code over $F_{q^m}^\eta$ with $d_R \geq d$, expand the elements of its parity-check matrix $H$ over the base field $\mathbb F_q$, and add appropriate parity-check equations to $H$ in $\mathbb F_q$ to ensure that the resulting code fits $S(v)$. 
\section{Ferrers Diagram Lifted Rank-Metric Codes for the injection Metric}\label{const}
Inspired by the construction of \cite{ES}, in this section we present a scheme for constructing $(n,M,d)_{d_I}$ Ferrers diagram lifted rank-metric codes in $\mathcal P_q^n$. The following theorem is key in our construction.
\begin{thm}\label{dma}
Let $U$ and $V$ be two vector spaces in $\mathcal P_q^n$, with profile vectors $u$, and $v$ respectively. Then we have, $d_I(U,V) \geq \max\{N(u,v),N(v,u)\} \nonumber$.
\end{thm}
\begin{proof}
First note that the dimension of a vector space is equal to the Hamming weight of its profile vector, i.e. $\dim U = wt(u)$ and $\dim V = wt(v)$. Now let $w= u \wedge v$ and observe that $\dim U \cap V \leq wt(w)$. Therefore we have $\dim U - \dim (U \cap V) \geq wt(u) - wt(w)$. Similarly, $\dim V - \dim (U \cap V) \geq wt(v) - wt(w)$. Taking the $\max$ of both equations we obtain, $d_I(U,V) \geq \max \{wt(u),wt(v)\} - wt(w) = \max\{N(u,v),N(v,u)\}$.
\end{proof}
For two binary vectors $x$ and $y$, the quantity $\max\{N(x,y),N(y,x)\}$ is a metric, known as the asymmetric distance between $x$ and $y$. The asymmetric distance was first introduced by Varshamov in \cite{asymmetric} for construction of codes for the Z channel. Constructions exist mainly for single-asymmetric error-correcting codes, and some multi-error correcting codes (\cite{klove} and references therein). Please refer to \cite{tasymm} for a more recent work on general $t$-asymmetric error-correcting codes.

By Theorem~\ref{dma} two spaces are guaranteed to have an injection distance of at least $d$, provided that the asymmetric distance between their profile vectors is at least $d$. Thus to construct a code in $\mathcal P_q^n$ with minimum injection distance $d$, we may select a set of subspaces according to an asymmetric code in the Hamming space with minimum asymmetric distance $d$ and follow a procedure similar to that presented in \cite{ES}. Construction of our $(n,M,d)_{d_I}$ code can be described algorithmically as follows:
\begin{enumerate}
\item Take a binary asymmetric code $\mathcal A$ of length $n$ and minimum asymmetric distance $d$. \label{step1}
\item For each codeword $c \in \mathcal A$, obtain $S(c)$, (composed of the columns of $P_M(c)$ with at least one $\bullet$).
\item Given each $k \times \eta$ matrix $S(c)$, use the construction of Section~\ref{fdrm} to obtain an $[S(c),\kappa,d]$ Ferrers diagram rank-metric code. \label{step3}
\item Lift each matrix $S(c)$ to its corresponding profile matrix $P_M(c)$, to obtain a generator matrix $G_c$. 
\item Finally $C = \{V \in \mathcal P_q^n| V = \left<G_c\right>\}$.
\end{enumerate}
Note that a slight modification to Step~\ref{step1} and Step~\ref{step3} in the above procedure allows for the construction of an $(n,M,d)_{d_S}$ code in the projective space. More specifically, in order to construct an $(n,M,2\delta)_{d_S}$ code in $\mathcal P_q^n$, we may first take a binary code $\mathcal H$ of minimum \textit{Hamming} distance $d_H \geq 2\delta$. Then for each codeword $c \in \mathcal H$ we may construct an $[S(c),\kappa,\delta]$ Ferrers diagram rank-metric code. Following the rest of the steps are described above, we obtain an $(n,M,2\delta)_{d_S}$ code. 
\section{Profile Vector Selection} \label{alg}
As suggested by Theorem~\ref{kappa}, the dimension of an $[S,\kappa,\delta]$ Ferrers diagram rank-metric code depends not only on the desired minimum distance $\delta$, but also on the number of $\bullet$'s in $S$. Since the number of $\bullet$'s in $S$ is directly related to its corresponding profile vector, the choice of the asymmetric code in the first step is crucial to the size of our codes. For instance, the vector $v=(1,1,0,0,0)$ results in a profile matrix with a higher number of $\bullet$'s than that of $v=(1,1,0,1,1)$. Thus a code of lower rate that contains vectors which potentially result in larger number of $\bullet$'s in their corresponding profile matrices may be preferable over one with a higher rate, that involves vectors resulting in smaller number of $\bullet$'s. 

With this observation, given a minimum asymmetric distance $d$ we define a scoring function $\score(v,d)$ on the set of all binary vectors, which calculates for every $v \in \{0,1\}^n$, the lower bound $\kappa$ of the dimension of the largest $[S(v), \kappa, d]$ Ferrers diagram rank-metric code induced by $v$. It is easy to observe that \begin{eqnarray}
\score(v,d) &=& \displaystyle \sum_{i=1}^n\sum_{j=1}^i \bar v_i v_j  -\max\{wt(v),\eta(v)\}(d-1)\nonumber \\
\mbox{where}~ \eta(v) &=& n-(wt(v)+\displaystyle \min_{t \in \operatorname{supp}(v)} t)+1 \nonumber
\end{eqnarray}

Now in order to select a set $P$ of profile vectors at a minimum asymmetric distance $d$, we use a standard greedy algorithm that maintains a list of available profile vectors $A \subseteq \{0,1\}^n$, (with $A$ initialized to $\{0,1\}^n$). At each step an available profile vector $v \in A$ with the largest score $\score(v,d)$ is added to $P$, and vectors within asymmetric distance $d$ of $v$ are made unavailable. The algorithm proceeds until $A = \emptyset$. By a slight modification to this algorithm we may allow for the same greedy selection of a set of profile vectors at a certain minimum \textit{Hamming} distance as opposed to a minimum \textit{asymmetric} distance.
%\begin{algorithm}
%\renewcommand{\columnwidth}{7cm}
%{\footnotesize
%\caption{Select a Set of Profile Vectors with $d_a \geq d$}
%\label{alg}
%\begin{algorithmic}
%\STATE $\Omega \leftarrow \{0,1\}^n$
%\STATE $\mathcal A \leftarrow \emptyset$
%\FOR{$k=1$ to $2^n$}
 %\STATE $x \leftarrow \displaystyle \argmax_{v \in (\Omega - \mathcal A)}\displaystyle \sum_{i=1}^n\sum_{j=1}^i \bar v_i v_j$
% \STATE $x \leftarrow \displaystyle \argmax_{v \in (\Omega - \mathcal A)}f(v)$
% {\STATE \tiny \begin{eqnarray} \mbox {where}~f(v) = \displaystyle \sum_{i=1}^n\sum_{j=1}^i \bar v_i v_j  -\max\{wt(v),n-wt(v)+1-\displaystyle \min_{t \in \operatorname{supp}(v)} t\}(d-1) \nonumber \end{eqnarray}}
 
%\IF{$(\mathcal A = \emptyset $ OR $\forall y \in \mathcal A \; d_a(x,y) \geq d)$}
%\STATE $c \leftarrow x$
%\ENDIF
%\STATE $\mathcal A \leftarrow \mathcal A \cup \{c\}$
%\STATE $k \leftarrow k+1$
%\ENDFOR
%\RETURN $\mathcal A$
%\end{algorithmic}}
%\end{algorithm}%

\section{Numerical Results}\label{params}
As constant-dimension codes designed for $d_S$ coincide with those designed for $d_I$, we are interested in the analysis of non-constant dimension $(n,M,d)_{d_I}$ codes.
\subsection{Our $(n,M,d)_{d_I}$ vs. $(n,M,d)_{d_S}$ Codes}
For our $(n,M,d)_{d_I}$ codes we first used the selection algorithm presented in Section~\ref{alg} to obtain a set of binary profile vectors at a minimum asymmetric distance $d_a \geq d$. Using this procedure along with the bound of Theorem~\ref{kappa} we obtained $\left|(n,M,d)_{d_I}\right|$. As discussed previously, for every $(n,M,d)_{d_I}$ code constructed according to the procedure described in Section~\ref{const}, we may construct an $(n,M,2d)_{d_S}$ counterpart through a similar procedure. In order to select a set of profile vectors for our $(n,M,2d)_{d_S}$ codes we used the algorithm of Section~\ref{alg} for $d_H \geq 2d$. As shown in Table~\ref{tb1} our $(n,M,d)_{d_I}$ codes denoted by $C_2$ have a slightly higher rate than their $(n,M,2d)_{d_S}$ counterparts, $C_1$.

\subsection{Our$(n,M,d)_{d_I}$ Codes vs. Codes of \cite{ES}}
The best $(n,M,d,k)_{d_S}$ constant-dimension codes of \cite{ES} are obtained by using constant-weight lexicodes as profile vectors. These codes achieve maximum cardinality when $k = \left \lfloor \displaystyle{\frac{n}{2}}\right \rfloor$. The column corresponding to $C_3$ in Table~\ref{tb1} shows rates of the $(n,M,d_S,\left \lfloor {\frac{n}{2}}\right \rfloor)_{d_S}$ constant-dimension codes of \cite{ES}. 

Non-constant-dimension $(n,M,d)_{d_S}$ codes of \cite{ES} are constructed by means of a puncturing operation performed on constant-dimension codes. As shown in \cite{ES} puncturing an $(n,M,d,k)_{d_S}$ code results in an $(n-1,M',d-1)_{d_S}$ code with $M' \geq \frac{M(q^{n-k}+q^k-2)}{q^n-1}$. In Table~\ref{tb1}, $\log_q\left|C_4\right|$ denotes the guaranteed minimum rate of $(n,M',d_S)$ punctured codes obtained from the best $(n+1,M,d_S+1,\left \lfloor {\frac{n+1}{2}}\right \rfloor)_{d_S}$ codes of \cite{ES}. As shown in the table, our $(n,M,d)_{d_I}$ codes have a slightly higher rate than both constant and non-constant-dimension codes of \cite{ES}.

\begin{table}
\caption{Parameters of codes constructed with $C_1 = \mbox{ our } (n,M,d_S)_{d_S}$ codes, $C_2 = \mbox{ our } (n,M,d_I)_{d_I}$ codes,$C_3 =(n,M,d_S,n/2)_{d_S}$ codes of \cite{ES},$C_4 = $ punctured codes of \cite{ES}}
 \centerline{\begin{tabular}{rrrrrrrrr}
$q$&$d_I$&$d_S$&$n$& $\log_q\left | C_1 \right |$ & $\log_q \left |C_2\right |$&$\log_q\left | C_3\right|$ & \multicolumn{1}{c}{$\log_q|C_4|$}\\
\hline
$2$&$2$&$4$&$9$ &$15.1732$&$15.6245$&$15.1731$&$10.9588$\\
$2$&$2$&$4$&$10$&$20.1551$&$20.3294$&$20.1548$&$13.5585$\\
$2$&$2$&$4$&$12$&$30.1561$&$30.3346$&$30.1559$&$13.7676$\\
$2$&$3$&$6$&$10$&$15.0031$&$15.0071$&$15.0032$&$7.5581$\\
$2$&$3$&$6$&$13$&$28.0032$&$28.0263$&$28.0032$&$21.9888$\\
$3$&$2$&$4$&$7$ &$8.0177$&$8.1331$&$8.0170$&$4.6210$\\
$3$&$2$&$4$&$8$ &$12.0138$&$12.0311$&$12.0138$&$6.2567$\\
$4$&$2$&$4$&$7$ &$8.0039$&$8.0522$&$8.0038$&$4.4974$\\
$4$&$2$&$4$&$8$ &$12.0031$&$12.0068$&$12.0031$&$6.1599$\\
\hline
\end{tabular}
\label{tb1}}
\end{table}
\section{Conclusion}
We presented a construction for the Ferrers diagram rank-metric codes as subcodes of linear MRD codes, and provided a lower bound on the dimension of the largest such codes. Using a greedy profile vector selection algorithm along with our construction of Ferrers diagram rank-metric codes we presented a class of non-constant dimension lifted Ferrers diagram rank-metric codes for the injection distance. We also presented a similar construction for non-constant dimension codes designed for the subspace distance. We observed that our non-constant dimension codes designed for the injection distance have a slightly higher rate than their counterparts designed for the subspace distance. Moreover, comparing our codes designed for the injection distance, with the best subspace codes of \cite{ES}, we observed a minor improvement in rate. The Ferrers diagram lifted rank-metric codes introduced by \cite{ES}, as well as those presented in our paper achieve higher rates than the original lifted rank-metric codes of \cite{KK}. However, we believe that these rate improvements are minute from a practical perspective. 
\bibliographystyle{IEEEtran}
\bibliography{main}
\end{document}